\documentclass[preprint, superscriptaddress, groupedaddress,amsmath,amssymb,aps,pre]{revtex4-1}

\usepackage{graphicx}
\usepackage{bm}
\usepackage[utf8]{inputenc}

\begin{document}

\title{Solving equations of motion by using Monte Carlo Metropolis: Novel method via Random Paths and Maximum Caliber Principle}

\author{Diego González$^{1,2}$, Sergio Davis$^{3}$, Sergio Curilef$^{1}$.}

\affiliation{$^{1}$Departamento de F\'\i sica, Universidad Católica del Norte, Antofagasta, Chile.}
\affiliation{$^{2}$Banco Itaú-Corpbanca, Region Metropolitana, Chile.}
\affiliation{$^{3}$Comisión Chilena de Energía Nuclear, Casilla 188-D, Santiago, Chile}
\date{\today}

\begin{abstract}
A permanent challenge in physics and other disciplines is to solve partial differential equations, thereby a beneficial investigation is to continue searching for new procedures to do it. In this Letter, a novel Monte-Carlo Metropolis framework is presented for solving the equations of motion in Lagrangian systems. The implementation lies in sampling the
paths space with a probability functional obtained by using the maximum caliber principle. The methodology was applied to the free particle and the harmonic oscillator problems, where the numerically-averaged path obtained from the Monte-Carlo simulation converges to the analytical solution from classical mechanics, in an analogous way with a canonical
system where energy is minimized by sampling the state space and computing the average state for each system. Thus, we expect that this procedure can be general enough to solve other differential equations in physics and to be a useful tool to calculate the time-dependent properties of dynamical systems in order to understand the non-equilibrium behavior of
statistical mechanical systems.

\end{abstract}

\maketitle


\section{Introduction}

The main objective of this work is to show a new framework for the study of dynamical systems which are described by a Lagrangian, being a first approach for the understanding of non-equilibrium statistical mechanics (NESM) by using constraints, as performed in statistical mechanics.

Here we propose a technique capable of simulating deterministic, dynamical systems through a stochastic formulation. This technique is based on sampling a statistical ensemble of paths defined by having the maximum path entropy (also known as the \emph{caliber}) available under imposed time-dependent constraints. This approach is known as the Maximum Caliber principle~\cite{Jaynes1980}, a generalization of Jaynes' principle of maximum entropy~\cite{Jaynes1957, Cafaro2016, General2018}.

The principle of Maximum Entropy (MaxEnt) is a systematic method for constructing the simplest, most unbiased probability distribution function under given constraints, a conceptual generalization of Gibbs' method of ensembles in Statistical Mechanics. The complete generality of the principle makes it widely used in several areas of Science, such as astronomy, ecology, biology, quantitative finance, image processing, electronics and physics among others. According to Jaynes, choosing a candidate probability distribution by maximizing its entropy is a rule of inferential reasoning far beyond its original application in Physics, which makes this rule a powerful tool for creating models in any context.

MonteCarlo Metropolis (MCM) is an computational algorithm for obtaining random samples drawn from a probability distribution. This probability distribution is usually constructed by using MaxEnt. MCM is used for the understanding of different systems in Physics \cite{Binder86,Graham2013} such as Spin Models, Material Simulations, Termodynamics, among others.
An analogous methodology is presented, by assigning probabilities to paths instead of probabilities of states, allowing the minimization of functional quantities such as the classical action instead of the energy.

It is possible to directly use MaxEnt instead of MaxCal as an inference methodology for dynamical systems, by constraining time-dependent functions, and this has been used to understand Newtonian dynamics~\cite{Caticha2007} and the Schr\"odinger equation~\cite{Caticha2011} in terms of traditional MaxEnt. A new approach has been explored recently for
recovering frameworks for dynamical systems by using MaxCal and the paths space ~\cite{Gonzalez2014, Davis2015, Gonzalez2016, Gonzalez2016b}, and this work shows a numerical implementation of this new approach, exposing the capability for solving complex problems in NESM. In summary, this constitutes a novel approach to the study of dynamical systems by using Maximum Caliber for Monte Carlo simulation.

\section{Creating a paths ensemble: The Maximum Caliber Principle}
\label{sec_maxcal}

The Maximum Caliber principle allows the definition of a unique paths ensemble given prior information and a number of dynamical constraints\cite{Jaynes1980, Grandy2008, Gonzalez2016}.

MaxCal is similar to the Maximum Entropy Principle (MaxEnt)\cite{Jaynes1982, Stock2008,Presse2013,Davis2015,Gonzalez2016b} but defined over the paths space, allowing to define a probability functional $P[x]$ as follows. Consider a path $x \in \mathbb{X}$. In order to construct a probability functional for each path $P[x]$, given an initial probability (prior) $P_0[x]$ and
an arbitrary constraint
\begin{equation}
\Big<A[x]\Big> = a,
\label{action}
\end{equation}
the Caliber (or path entropy)
\begin{equation}
S[P_0 \rightarrow P] = -\int_{\mathbb{X}} D_x P[x] \ln \frac{P[x]}{P_0[x]},
\end{equation}
must be maximized under the constraint in Eq. \ref{action} and the requirement that probability is normalized, $$\int_{\mathbb{X}} D_x P[x] = 1.$$
Then, the probability functional obtained is,
\begin{equation}
P[x|\beta]= \frac{1}{Z(\beta)}P_0[x]\exp(-\beta A[x]),
\label{probability}
\end{equation}
where $Z(\beta)$ is the partition function and $\beta$ is the Lagrange multiplier, given by the constraint equation
\begin{equation}
-\frac{\partial}{\partial \beta}\ln Z(\beta) = a.  
\end{equation}

\noindent
In this formalism, $\beta$ is analogous to the inverse of the temperature $\beta=1/k_B T$ in the canonical ensemble of Statistical Mechanics. Here the expected value of an arbitrary functional $F[x]$ is given by
\begin{equation}
\Big< F[x] \Big>_\beta = \frac{1}{Z(\beta)} \int_\mathbb{X} D_x \exp(-\beta A[x]) \; F[x],
\end{equation}
%
but, perhaps more importantly, the expectation value of a function over time can be defined similarly as
\begin{equation}
\Big< f(\dot x, x, t) \Big>_{\beta, t} = \frac{1}{Z(\beta)} \int_\mathbb{X} D_x \exp(-\beta A[x]) \; f(\dot x, x, t).
\end{equation}
%

This last relation shows that MaxCal can be used to understand macroscopic properties of time-dependent systems, which are the main elements in NESM.

\section{Least Action Principle and Most Probable Path}
\label{leastaction}

From classical mechanics it is well known that the path followed by a particle under a potential $V(x; t)$ under the boundary conditions $x(0)=0$ and $x(T)=x_f$ is the one given by the
least action principle~\cite{Lanczos1970, Feynman2005}, which in practice leads to an equation of motion describing the evolution of the particle from $0$ to $T$.

The classical action is a functional defined as $A[x]= \int_{0}^{T} dt \; L(\dot x,x;t)$, where $L(\dot x,x;t)$ is called the Lagrangian of the system, which for classical systems is
$$L(\dot x,x;t) = \frac{m \dot x^{2}}{2} - V(x;t).$$
For a MaxCal framework where the classical action is constrained, following an analogous treatment to the constraint in Eq. \ref{action} the probability functional is of the form given
in Eq. \ref{probability}. The most probable path can be obtained by finding the extrema of the functional $P[x]$, by solving the equation $\frac{\delta P[x]}{\delta x(t')} = 0$. This is
because the exponential function in Eq. \ref{probability} is convex and monotonically increasing, and so maximum probability is equivalent to imposing that $x$ should be an extremum for
the argument of the exponential,
\begin{equation}
- \beta \; \frac{\delta A[x]}{\delta x(t')} = 0.
\label{eq_extremum_A}
\end{equation}

If the Lagrange multiplier is positive ($\beta > 0$), the requirement is that the action is actually a minimum. This equation is precisely the Euler-Lagrange equation of motion for the Lagrangian~\cite{Lanczos1970,Gelfand2000}.
In summary, the most probable path and the least action path in a MaxCal framework are the same. According to this, sampling trajectories from the probability distribution in Eq. \ref{probability} using Monte Carlo methods and computing the averages of quantities, should converge to a description of the dynamical properties of a classical system evolving in time.

Another important consequence of the use of MaxCal and the form of the probability is that the most probable path in general coincides with the average path according to the central limit theorem.

\section{Computational method}

In order to define the elements on the paths space $\mathbb{X}$, for an $N$-dimensional path $x$, it is always possible to write it in a orthonormal basis $\phi_{i}$ of the form
\begin{equation}
x(t) = \sum_i^{n} a_i \phi_i(t) = x(t; \bm{a}).
\label{eq_basis_expand}
\end{equation}

Then, by changing the parameters $a_i$ it is possible to map the entire space of paths $x$. In other words, there is a one-to-one correspondence between an arbitrary path $x$ and its parameter vector $\bm{a}$, so the action becomes a function of $\bm{a}$, namely $\mathcal{A}(\bm{a}):= A[x(\cdot\; ;\bm{a})]$ and the problem of path sampling reduces to ordinary sampling
of $N$-dimensional states $\bm{a}$,
\begin{equation}
P(\bm{a}|\lambda) = \frac{1}{Z(\lambda)}\exp(-\lambda \mathcal{A}(\bm{a})).
\end{equation}

The choice of the basis functions $\phi_i$ is, in principle, arbitrary. However, a convenient choice can be made related to the particular conditions of the problem to be solved.
In this case, the target is the study of classical dynamical systems by using the MCM where the most of the problems have well-defined boundary conditions, therefore it is important to
find a basis set in which one can easily generate paths in the desired paths space holding the required, fixed boundary conditions. For this reason we considered the Bézier curves as
a basis set.

Bézier curves are defined by ``control points'' $c_{i}$, where the first $c_{0}$ and the last $c_{n}$ control point determine the boundary conditions of the curve, allowing the mapping
of the paths space with well-defined boundary conditions.

For an $N$-dimensional path $x$ with boundary conditions $x(t_{0})= c_{0}$ and $x(t_{f}) = c_{n}$, a Bézier curve is defined of the form
\begin{equation}
x(t) = \sum_{i=0}^{n} c_{i} \; B_i(t; n),
\label{path_bezier}
\end{equation}
where the basis functions $B_i(t; n)$ are the Bernstein polynomials,
\begin{equation}
B_{i}(t; n) = {n \choose i} \frac{(t-t_{0})^{i} (t_{f}-t)^{n-i}}{(t_{f}-t_{0})^{n}}.
\end{equation}

Following these definitions it is clear that all Bézier curves automatically follow the specified boundary conditions at $t=t_0$ and $t=t_f$.

\section{Implementation of the MonteCarlo Metropolis for sampling paths space}

The Monte Carlo Metropolis (MCM) implementation is usually employed for sampling a multidimensional space governed by a probability distribution~\cite{}. The following MCM
implementation is used for sampling the paths space $\mathbb{X}$ governed by a probability functional obtained via MaxCal. This procedure makes it possible to find the minimum action
path.

\begin{figure}
    \includegraphics[width=\textwidth]{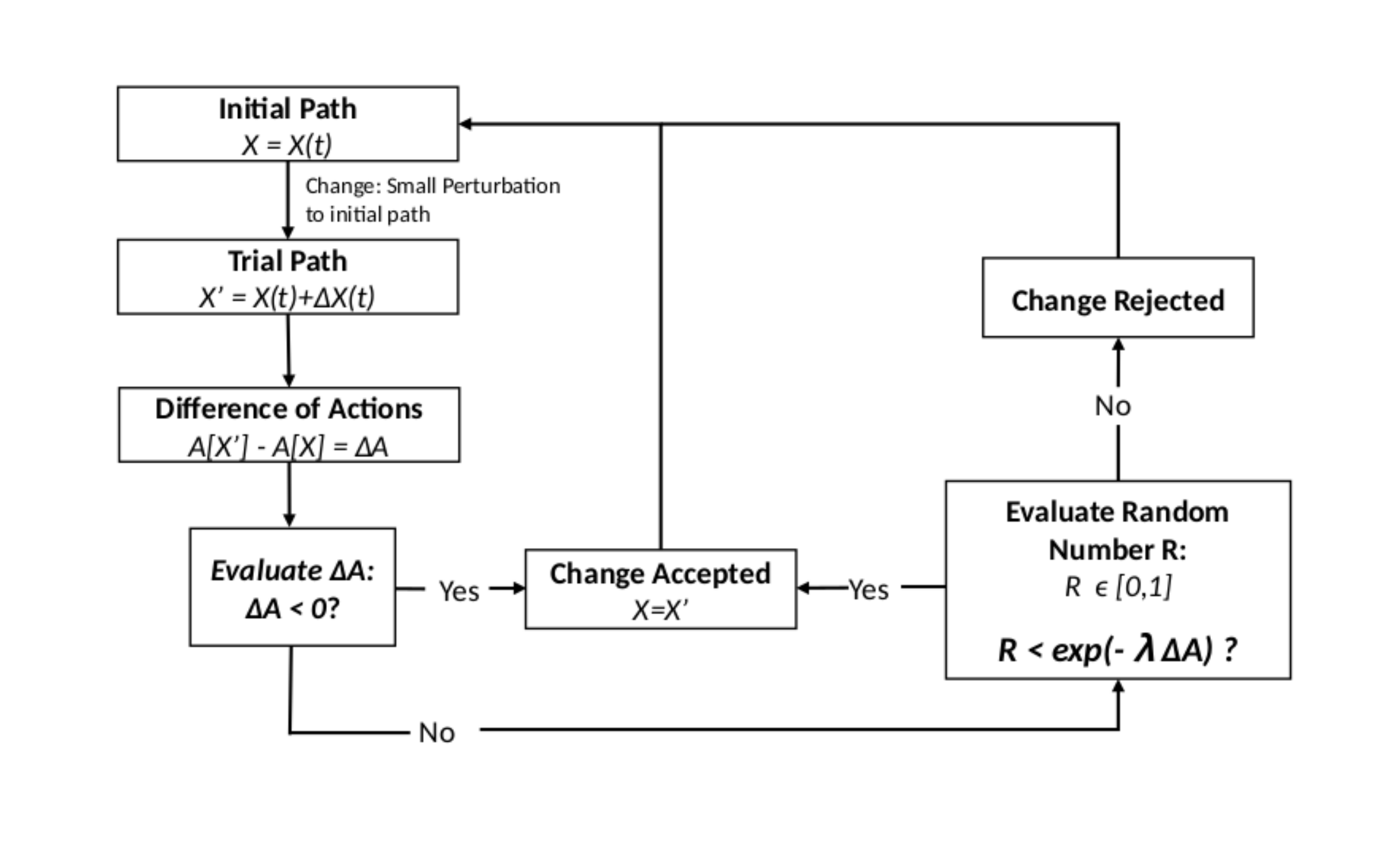}
    \caption{Explanatory diagram for a MCM sampling in paths space.}
    \label{MCMdiagram}
\end{figure}

For a given action $A[x]$ and boundary conditions $x(t_{0})= c_{0}$ and $x(t_{f}) = c_{n}$, the MCM evolution for samplig the paths space it is implemented as shown in Fig. \ref{MCMdiagram} 

By performing this process in an iterative way the paths space is sampled, allowing the calculation of properties for the system which is determined by the classical action used.
As shown in Section \ref{leastaction}, the probability distribution obtained when the classical action is constrained allows the sampling of the path space where the most probable
path and the least action path coincide. Finally, $\lambda$ can be related to the inverse of temperature as the usual MCM, due to the fact that, as $\lambda \rightarrow 0$ the sampled
paths will be random over the space, while taking $\lambda \rightarrow \infty$ constrains the sampled paths closer to the least action path. The value for $\lambda$ in a MCM is
related with the change from $x$ to $x'$, and empirically this change must be adjusted to have approximately a $80\%$ acceptance rate.

\section{Results and discussion}
\label{resultados}

\subsection{Free Particle Action}

The equation of motion for a free particle is obtained by minimizing the classical action $$A[x] = \int_{t_0}^{t_f} dt \; \frac{m \dot x(t)^2}{2}$$
according to the least action principle. For a free particle with mass $m=1$ and boundary conditions $x(0)=0$ and $x(1)=1$, the analytical solution for the least action path
is the straight line $x(t) = t$.

\begin{figure}
    \includegraphics[width=0.8\textwidth]{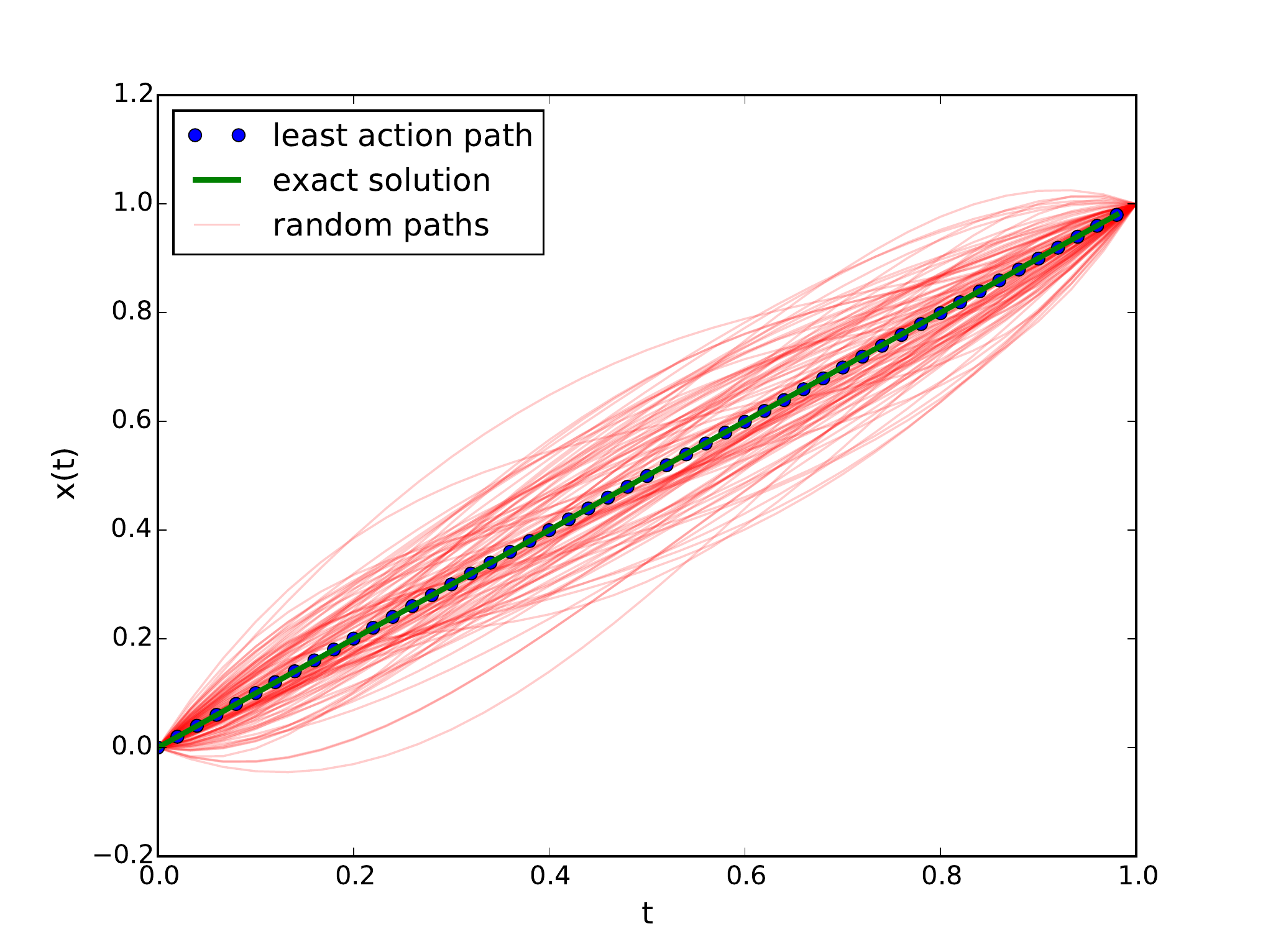}
    \caption{Dynamical trajectories sampled for the free particle, with boundary conditions \ldots.}
    \label{fig_freepart}
\end{figure}

By using MCM simulation as shown in Fig. \ref{fig_freepart}, we have sampled the path space and calculated the simple average position $\bar x(t)$ at each time. We see that
the simulation readily converges to the correct least action path.

For this simulation, 5 control points were used, and are sufficient to obtain the exact result ($R^{2}=0.99$) in less than 10 000 Monte Carlo steps, corresponding to
$\approx$ 18 min.

\subsection{Harmonic Oscillator Action}

In the case of the harmonic oscillator, the action is of the form
\begin{equation}
A[x] = \int_{t_{0}}^{t_{f}} {dt} \; \left(\frac{m \dot x(t)^{2}}{2} - \frac{k x(t)^{2}}{2}\right).
\end{equation}

Without loss of generality, for numerical simulations $m = 1$ and $k = 1$ are used.
The solution for this problem will be divided into two parts. The first solution found was the least action path for a short time interval. More precisely, we simulated a
particle with boundary conditions $x(0)=0$ and $x(t_{f})= a \sin(\omega t_{f})$, with $t_{f}$ less than the half period $\frac{T}{2}$.

\begin{figure}
    \centering
    \includegraphics[width=\textwidth]{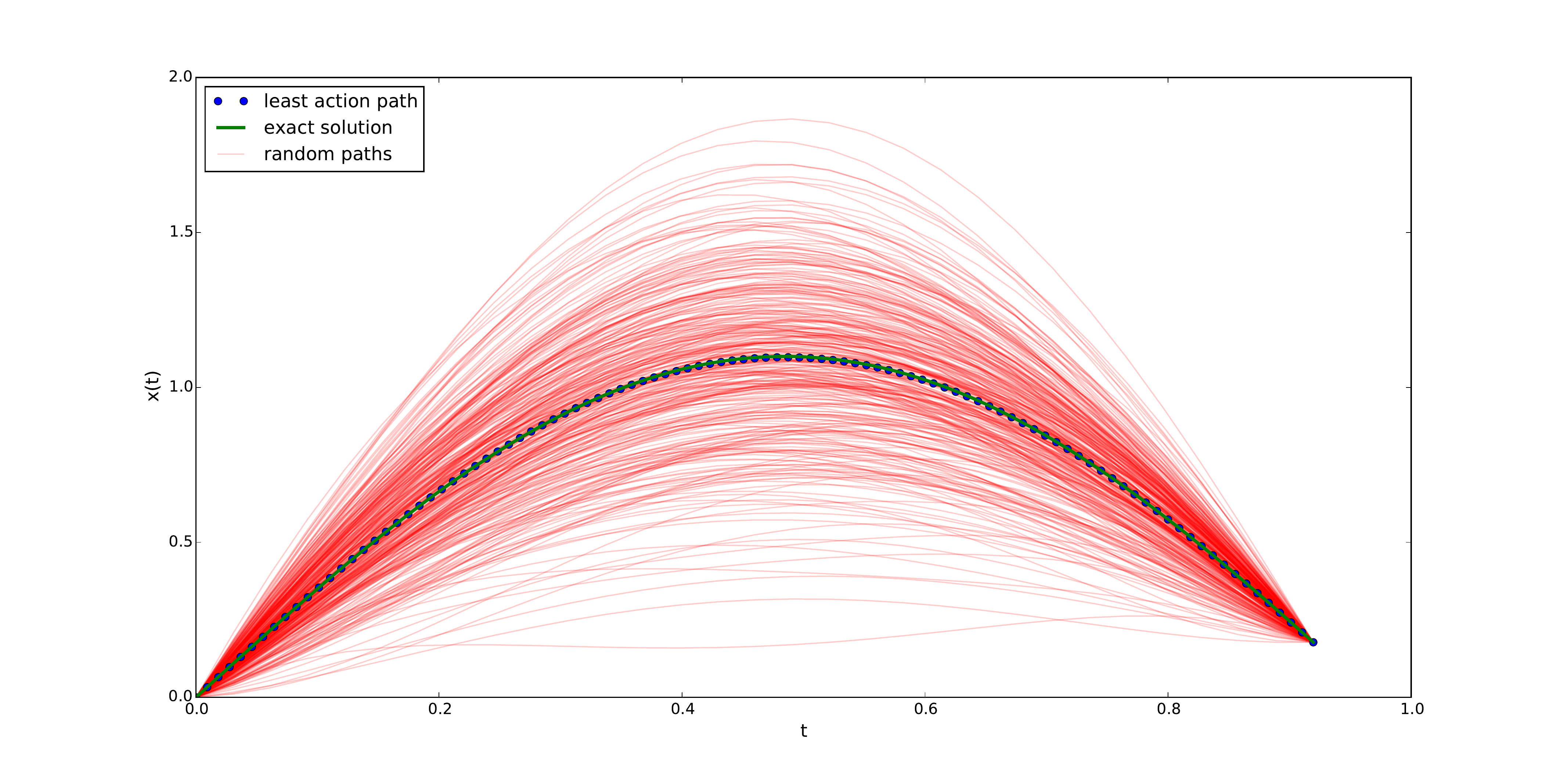}
    \caption{Paths sampled for the harmonic oscillator considering shorts time intervals.}
    \label{fig_HOpi}
\end{figure}

In this case the least action path also converges to the analytical solution, correctly solving the equation of motion as shown in Fig. \ref{fig_HOpi}. For this simulation, 5
control points were used, and are sufficient to obtain the exact result ($R^{2}=0.99$) in less than 10 000 Monte Carlo steps, corresponding to $\approx$ 26 min.

The second case corresponds to the harmonic oscillator with boundary conditions $x(0)=0$ and $x(t_{f})= a \sin(\omega t_{f})$, but where $t_{f}$ is larger than the half period
$\frac{T}{2}$. In other words, the end condition is past the first node of the harmonic oscillator. Under these boundary conditions, an unexpected result is obtained, the Monte Carlo
sampling procedure diverges. This result can be understood as due to the fact that, for this case, the action extremum is not a global minimum\cite{Gray2007}. More precisely, the
second functional derivative for the action of the harmonic oscillator action shows that the extremum is a saddle point in the case where the total time is longer than half a period,
and a ``true'' minimum only for paths with total time less than half period. We solved this convergence problem by considering an additional constraint to the action solved, suggested
by the work of Gray and Taylor\cite{Gray2007} in classical mechanics. The constraint involves the so-called kinetic foci, defined by the condition
\begin{equation}
\frac{\partial x}{\partial v_{0}} = 0.
\label{kineticfoci}
\end{equation}

As it turns out, the action extremum is guaranteed to be a minimum if the paths used pass close enough to a kinetic focus $x_{i}$. This can be implemented in the Monte Carlo simulation
by including a quadratic constraint in the probability functional, leading to
\begin{equation}
P[x|\lambda, \beta]= \frac{1}{\eta(\lambda, \beta)}P_0[x]\exp(-\lambda A[x] - \beta \sum_{i} (x(t_{i}) - x_{i})^{2}),
\label{probabilitykinecifoci}
\end{equation}
where $(t_{i}, x_{i})$ are the set of kinetic foci (solutions of Eq. \ref{kineticfoci}) and $\beta >> \lambda$, in order to stop the system from drifting away from the action extremum.

\begin{figure}
    \includegraphics[width=1.2\textwidth]{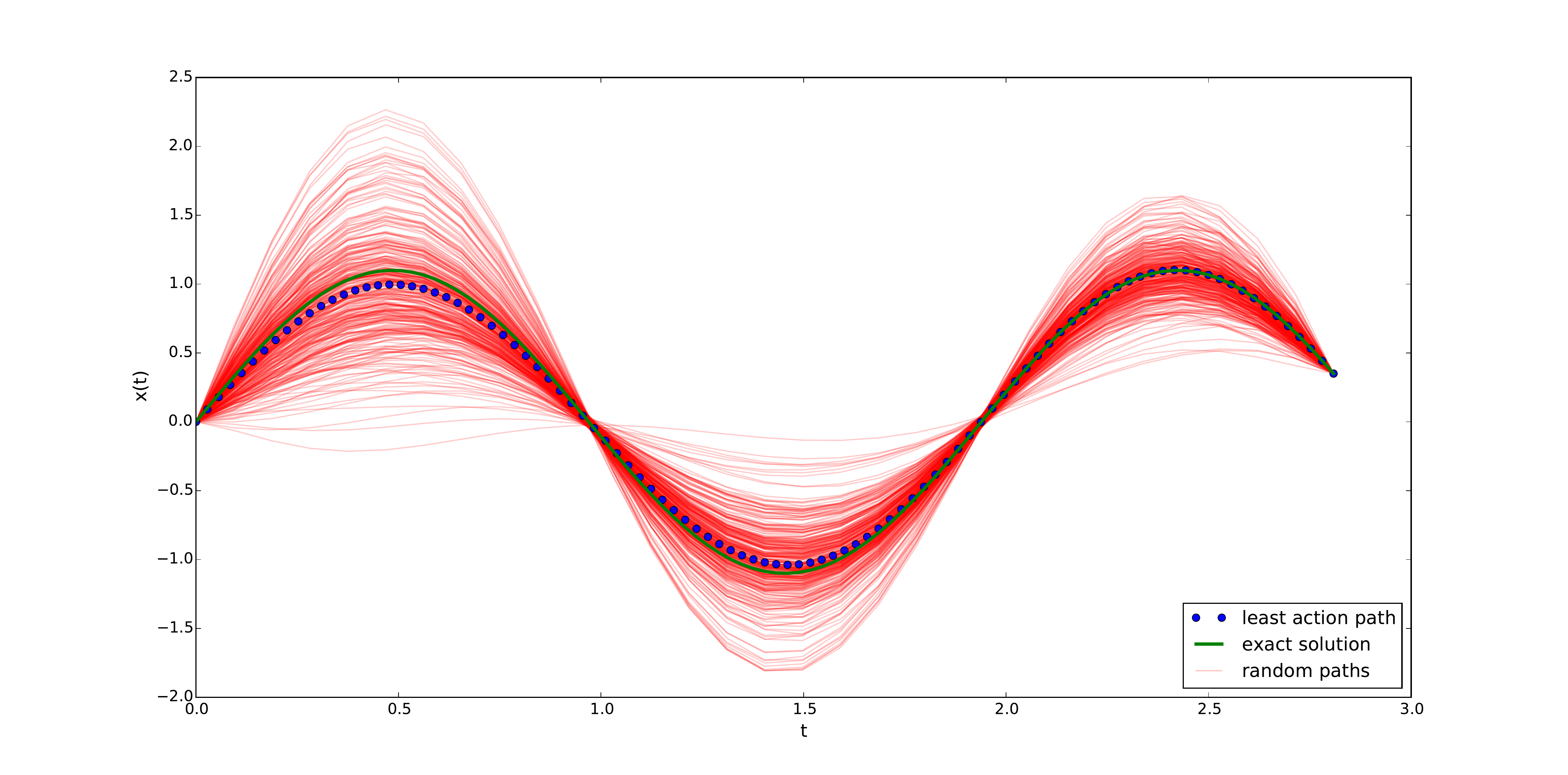}
    \caption{Harmonic oscillator with fixed kinetic foci.}
    \label{fig_HO3pi}
\end{figure}

As an example, we have solved the harmonic oscillator for a total time close to one and a half periods $\frac{3T}{2}$, sampling all the paths that cross the first two kinetic foci of
the harmonic oscillator, namely the points $\{ (0, \frac{T}{2}), (0,T)\}$. Now the Monte Carlo procedure does converge to the expected solution as shown in Fig. \ref{fig_HO3pi}.

For this simulation, 8 control points were used, and were sufficient to reach the exact result ($R^{2}=0.95$) in less than 20 000 Monte Carlo steps, corresponding to $\approx$ 2.8 hours.

An important remark here is to note the number of Bézier basis elements, or control points, because this is closely related to the computing time. For this reason, the main goal
in an efficient simulation is to find the minimum number of control points to use without sacrificing precision, needed to map any solution of the differential equation of interest.

\section{Concluding remarks}

In summary, we have described a technique for implementing Monte Carlo sampling of dynamical trajectories in classical Lagrangian systems under the Maximum Caliber formalism. We have demonstrated its usefulness by applying this technique to the case of the classical free particle and harmonic oscillator, recovering in both cases a statistical distribution of
paths centered on the classical solution of the Euler-Lagrange equation. For the case of the harmonic oscillator we noted the need for fixing additional points known as the kinetic foci of the system in order for the simulation to converge properly. Our proof-of-concept implementation could be the starting point for a complete computational scheme of simulation of
classical systems under uncertainty. It remains to be seen how this method scales to multidimensional and many-particle interacting systems. Finally, one of the main proposed uses of this
framework is to obtain the instantaneous probability density of positions at each time, which would allow to obtain the instantaneous macroscopic properties of classical systems under uncertainty\cite{Gonzalez2016c}.

\section*{Acknowledgments}
DG acknowledges funding from ``Beca de postdoctorado Universidad Católica del Norte'' N0 0004/2019, and Itaú-Corpbanca, and Dr. Gonzalo Gutiérrez for the support as tutor during the PhD. where this ideas started. SD acknowledges partial financial support from FONDECYT grant 1171127 and Anillo ACT-172101. SC acknowledges partial financial support from FONDECYT grant 1170834.

\bibliography{maxcal}
\bibliographystyle{apsrev}

\end{document}